\documentclass[11pt]{article}

\renewcommand{\baselinestretch}{1.2}
  \renewcommand{\arraystretch}{1.0}
\usepackage{cite}
 \usepackage{amsmath}
\usepackage{amscd}
\usepackage{amssymb}
\usepackage{mathtools}
\usepackage{graphicx}
\usepackage{multirow}
\usepackage{ifpdf}
\usepackage{fancyhdr}

\begin{document}

 \title{The Paillier's Cryptosystem and  Some Variants Revisited}
 \author{Zhengjun Cao$^{1}$, \qquad Lihua Liu$^{2,*}$}
  \footnotetext{ $^1$Department of Mathematics, Shanghai University, Shanghai,
  China. \quad      $^2$Department of Mathematics, Shanghai Maritime University,   Shanghai,
  China.   $^*$\,\textsf{liulh@shmtu.edu.cn}    }

\date{}
\maketitle

\begin{quotation}\noindent \textbf{Abstract}. At Eurocrypt'99, Paillier presented a public-key cryptosystem based on a novel computational problem.  It has interested many researchers because it was additively homomorphic. In this paper, we show that there is a big difference between the original Paillier's encryption and some variants. The Paillier's encryption  can be  naturally transformed into a signature scheme but these variants miss the feature.   In particular, we  simplify  the alternative decryption procedure of Bresson-Catalano-Pointcheval encryption scheme proposed at Asiacrypt'03.    The new version is more applicable to cloud computing because of its  double trapdoor decryption mechanism and its flexibility to be integrated into other cryptographic schemes. It captures a new feature that its two groups of secret keys can be distributed to different users so as to enhance the robustness of key management.

 \end{quotation}

\section{Introduction}
Homomorphic encryption is a very useful cryptographic primitive because it can translate
 an operation on the ciphertexts into an operation on the underlying plaintexts.
  The property is very important for many applications, such as e-voting, threshold cryptosystems, watermarking and secret sharing schemes. For example,  if an additively homomorphic
encryption is used in an e-voting scheme, one can obtain an encryption of the sum of all
ballots from their encryption. Consequently, it becomes possible that a single decryption will reveal the result
of the election. That is, it is unnecessary to decrypt all ciphertexts one by one.

At Eurocrypt'99, Paillier \cite{P99} proposed a public-key cryptosystem based on a novel computational problem.    It encrypts a message $m$ by
$$E(m, r) = g^mr^n \!\!\!\mod  n^2,$$ where $n=pq$ is an RSA modulus, $g$ is a public parameter such that $n\,|\,\mbox{ord}_{n^2}(g)$,  and $r$ is a random pad. The encryption function $E(m, r)$ has the additively homomorphic property, i.e.,
$$E(m_1,  r_1)E(m_2, r_2) = E(m_1 +m_2, r_1r_2).$$ More powerful, who knows the trapdoor of the encryption function can  recover not only the message $m$ but also the random pad $r$. This is another appreciated property for many applications.  Due to this property, the Paillier's encryption scheme can be naturally transformed into a one-way trapdoor permutation and a digital signature scheme.

\emph{Related Work.}  In 1984, Goldwasser and Micali \cite{GM84} proposed the first probabilistic
encryption scheme which was also homomorphic.   It has been improved in \cite{NS98,OU98}. In 1999, Paillier \cite{P99} presented a novel additively homomorphic encryption which was more powerful because it can recover the random pad $r$  as well as the message $m$.
At PKC'01, Damg{\aa}rd and Jurik \cite{DJ01} put forth a generalization of Paillier's encryption using
computations modulo $n^{i} (i\geq 2)$ and taking a special base $g=n+1$.
 They \cite{DJN10} also investigated the applications of the generalization. The elliptic curve variant of Paillier's cryptosystem is due to Galbraith \cite{G02}.

In 2001, Choi et al. \cite{CCW01} revisited the Paillier's encryption by taking a special base $g$ such that $g^{\lambda}=1+n\!\!\!\mod  n^2$, where
$\lambda=\mbox{lcm}(p-1, q-1)$.  Shortly after that, Sakurai and Takagi \cite{ST02} pointed out that the variant cannot resist  a chosen
ciphertext attack which  can factor the modulus $n$ by only one query to the decryption oracle.

At Eurocrypt'06, Schoenmakers and Tuyls \cite{ST06} have  considered
the problem of converting a given Paillier's encryption of a value $x \in \mathbb{Z}_n$
into Paillier's encryptions of the bits of $x$.  At Eurocrypt'13, Joye and Libert \cite{JL13} obtained another generalization based on $2^k$-th power residue problem. In 2013, Boneh et al. \cite{BCH13} considered the problem of  private database queries using Paillier's homomorphic encryption.  At Asiacrypt' 14, Catalano et al. \cite{CMP14} presented an instantiation of publicly verifiable delegation of computation on outsourced ciphertext which supports Paillier's encryption. In 2015, Castagnos and Laguillaumie \cite{CL15} designed a linearly homomorphic encryption scheme whose security relies on the hardness of the decisional Diffie-Hellman problem. Their scheme is somehow similar to the one of \cite{BCP03}.
The Gentry's fully homomorphic encryption scheme \cite{G09} relies on hard problems related to lattices, which actually allows to evaluate
any function on messages given their ciphertexts. But Paillier's cryptosystem based on the problem of factoring
RSA integers is still more competitive for applications that need only to add ciphertexts.

 \emph{Our contributions}.  In this paper, we revisit the Paillier's cryptosystem and reaffirm that the Paillier's encryption can be naturally transformed into a signature scheme but some variants miss the feature. Our presentation of the cryptosystem and some variants is so plain and heuristic that it becomes possible to investigate the further applications of these schemes in different scenarios.  In particular, we simplify the alternative decryption procedure of Bresson-Catalano-Pointcheval encryption scheme.  Our new proposal is more applicable to cloud computing because of its  double trapdoor decryption mechanism and its flexibility to be integrated into other cryptographic schemes.
It captures a new feature that its two groups of secret parameters can be allocated to different users so as to enhance the robustness of key management.

 \section{Paillier's encryption scheme}

Let $n = pq$ be an RSA modulus and $\phi(n)$ be the Euler's totient function. Set $\lambda = \mbox{lcm} (p-1, q-1)$.  Hence,  $|\mathbb{Z}^*_{n^2}|=\phi(n^2)=n\phi(n)$ and  for any $w \in \mathbb{Z}^*_{n^2}$
  $$w^{\lambda}=1\!\!\!\mod n, \quad  w^{n\lambda}=1 \!\!\!\mod  n^2 $$
  which are due to Carmichael's theorem.

  \textbf{Definition 1}. \emph{A number $z$ is said to be a $n$-th residue modulo $n^2$ if there exists
a number $y\in \mathbb{Z}^*_{n^2}$  such that $ z = y^n \!\!\!\mod  n^2.$ }

The set of $n$-th residues is a multiplicative subgroup of $\mathbb{Z}^*_{n^2}$ of order $\phi(n)$.
Each $n$-th residue  has exactly $n$ roots, among which exactly one
is strictly smaller than $n$.

  Let $g$ be some element of $\mathbb{Z}^*_{n^2}$
and define the following integer-valued function
$$ \mathcal{E}_g: \quad \mathbb{Z}_{n}\times \mathbb{Z}^*_{n}\longmapsto  \mathbb{Z}^*_{n^2} $$
$$ \qquad\qquad\quad (x, y) \longmapsto g^x\cdot y^n \!\!\!\mod  n^2. $$

\textbf{Lemma 1}. \emph{If  $n\,|\,\mbox{ord}_{n^2}(g)$, then $\mathcal{E}_g$ is bijective.}

\emph{Proof}.  Since the two groups $\mathbb{Z}_{n}\times \mathbb{Z}^*_{n}$
 and $\mathbb{Z}^*_{n^2}$ have the same number of elements $n\phi(n)$, it suffices to prove that $\mathcal{E}_g$ is injective.
 Suppose that $g^{x_1}y_1^{n} =g^{x_2}y_2^{n}  \!\!\!\mod  n^2$, where $x_1, x_2 \in \mathbb{Z}_{n}, y_1, y_2\in \mathbb{Z}^*_{n}$. It comes $g^{x_2-x_1} (y_2/y_1)^n = 1 \!\!\!\mod  n^2$, which implies $g^{\lambda(x_2-x_1)}(y_2/y_1)^{\lambda n}=g^{\lambda(x_2-x_1)}=1 \!\!\!\mod  n^2$. Thus $\mbox{ord}_{n^2}(g)\,|\,\lambda(x_2-x_1)$.
 Since $n\,|\,\mbox{ord}_{n^2}(g)$, we have $n\,|\,\lambda(x_2-x_1)$. In view of that $(n, \lambda)=1$, we obtain  $x_2=x_1 \!\!\!\mod  n$.  Since $x_1, x_2 \in \mathbb{Z}_{n}$, it comes $x_1=x_2.$  Thus, $(y_2/y_1)^n = 1 \!\!\!\mod  n^2$, which leads to the unique solution $y_2/y_1=1$  over $\mathbb{Z}^*_{n}$. This means $x_1=x_2$ and $y_1=y_2$. Therefore, $\mathcal{E}_g$ is bijective.  \hfill $\Box$

 By the above lemma, for a given $w\in  \mathbb{Z}^*_{n^2}$, there exists a pair $(x, y)$ such that $w=g^xy^n \!\!\!\mod  n^2.$

\textbf{Problem 1}.  \emph{Given an RSA modulus $n=pq$, $c, g \in  \mathbb{Z}^*_{n^2}$, \fbox{compute $x\in \mathbb{Z}^*_n$} such that
$$g^xy^n=c\!\!\!\mod  n^2,  $$
where  $n\,|\,\mbox{ord}_{n^2}(g)$  and $y$ is some element of $\mathbb{Z}^*_{n^2}$.}

\textbf{Theorem 1}. \emph{If $\lambda$ is known and $(\frac{g^{\lambda}-1 \!\!\!\mod  n^2}{n}, n)=1$, then one can solve Problem 1 by computing
$$x= \left(\frac{c^{\lambda}-1 \!\!\!\mod  n^2}{n}\right)  \left(\frac{g^{\lambda}-1 \!\!\!\mod  n^2}{n}\right)^{-1} \!\!\!\mod  n.  $$
}

\emph{Proof}. By the definition of $\lambda$, it comes  $c^{\lambda}=1 \!\!\!\mod  n, g^{\lambda}=1 \!\!\!\mod  n$. Set
$$ c^{\lambda} =a n+1\!\!\!\mod  n^2, \quad g^{\lambda} =b n+1\!\!\!\mod  n^2,$$
i.e., $$ a=\frac{c^{\lambda}-1 \!\!\!\mod  n^2}{n}, \quad  b=\frac{g^{\lambda}-1 \!\!\!\mod  n^2}{n}.$$
Since $n\,|\,\mbox{ord}_{n^2}(g)$,  $\mathcal{E}_g$ is bijective. There exists a pair $(x, y)\in \mathbb{Z}_{n}\times \mathbb{Z}^*_{n}$ such that $c=g^xy^n\!\!\!\mod  n^2$.
Hence,
$c^{\lambda}=(g^xy^n)^{\lambda} \!\!\!\mod  n^2. $
Since
 $y^{n\lambda}=1 \!\!\!\mod  n^2$, it comes
$ c^{\lambda}=(g^{\lambda})^x \!\!\!\mod  n^2.$
Thus,
$$a n+1=(b n+1)^x=xbn+1\!\!\!\mod  n^2 $$
this is due to $n^2\,|\,{x \choose i}(bn)^i, i\geq 2$.
Therefore,
$a n=xbn\!\!\!\mod  n^2$. That means
 $a =xb\!\!\!\mod  n.$
Since $(b, n)=1$, it gives $x=ab^{-1}\!\!\!\mod  n$.
\hfill $\Box$

\emph{Remark 1}. Paillier called the Problem 1 as Composite Residuosity Class Problem (see Definition 8 in Ref.\cite{P99}).  In view of that the trapdoor $\lambda$ plays a key role in computing the exponent $x$ with respect to the base $g$,
 we would like to call the Problem 1 as Trapdoored  Partial Discrete Logarithm Problem.

\textbf{Conjecture 1}.  \emph{If the trapdoor $\lambda$ is unknown, there exists no probabilistic polynomial time algorithm that
solves the Problem 1.}

Based on the above results, at Eurocrypt'99 Paillier proposed his cryptosystem.  The cryptosystem includes a probabilistic encryption scheme, a one-way trapdoor permutation and a digital signature scheme. We now describe the encryption scheme  as follows.

\begin{center}
Table 1: Paillier's encryption scheme

\begin{tabular}{|l|l|}
  \hline
  Setup & Pick an RSA modulus $n=pq$.  Set $\lambda = \mbox{lcm} (p-1, q-1)$.\\
 & Select $g\in \mathbb{Z}^*_{n^2}$ such that
 $n\,|\,\mbox{ord}_{n^2}(g)$.
 \\
 &  Publish $n, g$ and keep $\lambda$ in secret.
 \\  \hline
 Enc. & For $m\in \mathbb{Z}_{n}$, pick $r\in \mathbb{Z}_{n}$,  compute the ciphertext \\
 & \qquad\qquad\qquad $c=g^mr^n \!\!\!\mod  n^2.$
  \\ \hline
  Dec. & Recover $m=\left(\frac{c^{\lambda}-1 \!\!\!\mod  n^2}{n}\right)/\left(\frac{g^{\lambda}-1 \!\!\!\mod  n^2}{n}\right) \!\!\!\mod  n $\\
   \hline
\end{tabular}\end{center}

\section{A Hybrid Computational Problem}

We now consider another computational problem which is a hybrid of partial discrete logarithm problem  and $n$-th residuosity problem.

\textbf{Problem 2}.  \emph{Given an RSA modulus $n=pq$, $c, g \in  \mathbb{Z}^*_{n^2}$, \fbox{compute $(x, y)\in\mathbb{Z}_{n}\times \mathbb{Z}^*_{n}$} such that
$$g^xy^n=c\!\!\!\mod  n^2$$ where $n\,|\,\mbox{ord}_{n^2}(g)$.}

Notice that the solvability of Problem 2 directly implies that of Problem 1. We shall prove that the inverse holds, too.

\emph{If the trapdoor $\lambda$ is known}, Paillier proposed a method to solve the hybrid computational problem. He pointed out that $x, y$ can be computed by
$$ x= \left(\frac{c^{\lambda}-1 \!\!\!\mod  n^2}{n}\right)  \left(\frac{g^{\lambda}-1 \!\!\!\mod  n^2}{n}\right)^{-1} \!\!\!\mod  n, \quad y=(cg^{-x})^{1/n\!\!\!\mod\lambda}\!\!\!\mod  n.  $$

The idea behind his method can be described as follows.
By the existence of $(x, y)$, it is easy to find that
\begin{eqnarray*}
  g^xy^n=c\!\!\!\mod  n^2\ &\Longrightarrow& \  g^xy^n=c \!\!\!\mod  n
 \Longleftrightarrow   y^n=cg^{-x} \!\!\!\mod  n \\
 &\Longleftrightarrow&   (y^n)^{1/n\!\!\!\mod\lambda}=(cg^{-x})^{1/n\!\!\!\mod\lambda} \!\!\!\mod  n \\
  &\Longleftrightarrow&   y=(cg^{-x})^{1/n\!\!\!\mod\lambda} \!\!\!\mod  n
 \end{eqnarray*}
 By the uniqueness of $(x, y)\in \mathbb{Z}_{n}\times \mathbb{Z}^*_{n}$, we conclude that it is properly computed.

  \textbf{Theorem 2}. \emph{If $\lambda$ is known and $(\frac{g^{\lambda}-1 \!\!\!\mod  n^2}{n}, n)=1$, then one can solve Problem 2 by computing
$$x= \left(\frac{c^{\lambda}-1 \!\!\!\mod  n^2}{n}\right)  \left(\frac{g^{\lambda}-1 \!\!\!\mod  n^2}{n}\right)^{-1} \!\!\!\mod  n, \qquad y=(cg^{-x})^{s} \!\!\!\mod  n, $$
where $s$ is the integer with the least absolute value
such that $\lambda\,|\, ns-1$.
}

\emph{Proof}. Since $(n, \lambda)=1$, it is easy to compute the integer $s$ with the least absolute value  such that $\lambda\,|\, ns-1$.
By Theorem 1, we conclude that $x$ is properly  computed.
By the existence of $y$ and $y^n=cg^{-x}\!\!\!\mod  n^2$, we have
$$(cg^{-x})^{\lambda}=y^{n\lambda}=1\!\!\!\mod  n^2  $$
Now, suppose that $ns-1=\lambda\phi$  and $(cg^{-x})^{s}=\ell n+ y  \!\!\!\mod  n^2$ for some integers $\phi, \ell$.  Hence,  it comes
$$ g^x\left((cg^{-x})^{s}-\ell n\right)^n=g^x(cg^{-x})(cg^{-x})^{ns-1}=c(cg^{-x})^{\lambda\phi}=c((cg^{-x})^{\lambda})^{\phi}=c\cdot 1^{\phi}=c\quad \mbox{mod}\  n^2$$
 This completes the proof. \hfill $\Box$

 Note that the values  $s $ and  $\left(\frac{g^{\lambda}-1 \!\!\!\mod  n^2}{n}\right)^{-1} \!\!\!\mod  n$ have no relation to the ciphertext  $c$. They can be computed and stored previously.

 \textbf{Conjecture 2}.  \emph{If $\lambda$ is unknown, there exists no probabilistic polynomial time algorithm that
solves the Problem 2.}

\section{The Paillier's one-way trapdoor permutation and the digital signature scheme}

In Ref.\cite{P99}, Paillier has put forth  a one-way trapdoor permutation and the digital signature scheme based on his computational method. We now relate them as follows.

\begin{center}
Table 2: Paillier's signature scheme

\begin{tabular}{|l|l|}
  \hline
 Setup & See Table 1.
 \\  \hline
\multirow{4}*{\centering\parbox{0.5in}{Sign}}  & For a message $m$, compute   \\
 & $s_1\leftarrow \rho \left(\frac{H(m)^{\lambda}-1 \!\!\!\mod  n^2}{n}\right) \!\!\!\mod  n.$  \\
 & $s_2\leftarrow ((H(m)g^{-s_1})^s\!\!\!\mod  n$ \\
 & The signature is $(m; s_1, s_2)$.   \\ \hline
  Verify & $H(m)\stackrel{?}{=} g^{s_1}s_2^n \!\!\!\mod  n^2 $ \\
  \hline
\end{tabular}\end{center} \newpage

\begin{center}
Table 3: Paillier's one-way  trapdoor permutation

\begin{tabular}{|l|l|}
  \hline
 \multirow{3}*{\centering\parbox{0.5in}{Setup}} & Set an RSA modulus $n=pq$,  $\lambda = \mbox{lcm} (p-1, q-1)$. \\
 &    Select $g\in \mathbb{Z}^*_{n^2}$ such that $n\,|\,\mbox{ord}_{n^2}(g)$.   \\
&  Compute $\rho=\left(\frac{g^{\lambda}-1 \!\!\!\mod  n^2}{n}\right)^{-1} \mbox{mod}\  n,  $  and compute $s$ which is\\
&  the integer with the least absolute value such that $\lambda\,|\, ns-1$.\\
 & Publish $n, g$ and keep $\lambda, \rho, s$ in secret.
 \\  \hline
 \multirow{2}*{\centering\parbox{0.8in}{Encryption}}
  & Given  $m\in \mathbb{Z}_{n^2}$, set $m=m_1+nm_2$.   \\
  & The ciphertext is $c\leftarrow g^{m_1}m_2^n \!\!\!\mod  n^2.$
  \\ \hline
  \multirow{3}*{\centering\parbox{0.8in}{Decryption}} & $m_1\leftarrow \rho \left(\frac{c^{\lambda}-1 \!\!\!\mod  n^2}{n}\right) \!\!\!\mod n, $ \quad $m_2\leftarrow (cg^{-m_1})^s \!\!\!\mod  n.$\\
  &  $m\leftarrow m_1+nm_2.$  \\ \hline
\end{tabular}\end{center}\vspace*{3mm}

\begin{center}
Table 4: Some variants of Paillier's encryption schemes

\begin{tabular}{|l|l|}
  \hline
  &   $n=pq$ is an RSA modulus,  $\lambda=\mbox{lcm}(p-1, q-1)$.   \\ \hline
 Variant 1 &  $g\in \mathbb{Z}^*_{n^2}$, $\mbox{ord}_{n^2}(g)=\alpha n$. PK: $n, g$; SK:   $\alpha$.  \\  \cline{2-2}
  (Paillier) &  $m\in \mathbb{Z}_{n}$,  $r\in \mathbb{Z}_{n}$,   $c= g^{m+rn} \!\!\!\mod  n^2.$    \\ \cline{2-2}
& \qquad\qquad  $m= \left(\frac{c^{\alpha}-1 \!\!\!\mod  n^2}{n}\right)/\left(\frac{g^{\alpha}-1 \!\!\!\mod  n^2}{n}\right) \!\!\!\mod  n $  \\ \cline{2-2}
   \hline \hline
      Variant 2  &   $\kappa=\tau\lambda$, $\tau=\lambda^{-1} \!\!\!\mod  n$. PK: $n$; SK:   $\kappa$.  \\  \cline{2-2}
     (Damg{\aa}rd-Jurik)  &  $m\in \mathbb{Z}_{n}$,  $r\in \mathbb{Z}_{n}$,  $c=(1+mn) r^n\!\!\!\mod  n^2.$    \\ \cline{2-2}
& \qquad\qquad $m=\frac{c^{\kappa}-1 \!\!\!\mod  n^2}{n}$ \\ \cline{2-2}
\hline  \hline
  Variant 3    &  $g^{\lambda}=1+n\!\!\!\mod  n^2.$ PK: $n, g$; SK:   $\lambda$.  \\  \cline{2-2}
    (Choi-Choi-Won)  & $m\in \mathbb{Z}_{n}$, $r\in \mathbb{Z}_{n}$,  $c=g^m r^n \!\!\!\mod  n^2.$    \\ \cline{2-2}
 &\multirow{2}*{\centering\parbox{3.5in} {\qquad\qquad $m=\frac{c^{\lambda}-1 \!\!\!\mod  n^2}{n}$ }} \\
 & \\  \cline{2-2}
  \hline \hline
  Variant 4    &  $e<n$,  $d=e^{-1} \!\!\!\mod  \phi(n)$.  PK: $n, e$; SK:   $d$.  \\  \cline{2-2}
    (Catalano-Gennaro&  $m\in \mathbb{Z}_{n}$,  $r\in \mathbb{Z}_{n}$,  $c=(1+mn) r^e\!\!\!\mod  n^2.$    \\ \cline{2-2}
   -Howgrave-Nguyen) & \multirow{2}*{\centering\parbox{3.5in} {\qquad\qquad $m=\frac{\ \ \frac{c}{(c^d\!\!\!\mod  n)^e}-1\!\!\!\mod  n^2}{n}$}\  \ }  \\
     & \\  \cline{2-2}
    \hline
\end{tabular}\end{center}

\section{On some variants of Paillier's encryption scheme}

\subsection{Descriptons of some variants}

In the same article \cite{P99}, Paillier  has pointed out that there was an efficient variant of his original encryption scheme.
 Shortly afterwards, other variants came out \cite{BCP03,CGH01,CCW01,DJ01}. We list some variants as follows.

\textbf{Correctness of Variant 1}. The variant takes $x=m, y=g^r$ in Problem 2.
 Since each $n$-th residue has exactly $n$ roots, among which exactly one
is strictly smaller than $n$, and  $\mbox{ord}_{n^2}(g)=\alpha n$,  we have $g^{\alpha}=1 \!\!\!\mod  n$. Otherwise, suppose $g^{\alpha}=sn+t \!\!\!\mod  n^2$ for some integers $0\leq s<n$ and $t\,  (2\leq t<n)$. It leads to
$$1=g^{\alpha n}=(sn+t)^n=t^n  \!\!\!\mod  n^2 $$
 which means $t=1$.  It is a contradiction. Thus,  $g^{\alpha}=sn+1 \!\!\!\mod  n^2$. By
$$c^{\alpha}=(g^m(g^r)^n)^{\alpha}=(g^{\alpha})^m=(sn+1)^m=smn+1\!\!\!\mod  n^2$$
 we have
 $$\frac{\ \ \ \frac{c^{\alpha}-1 \!\!\!\mod  n^2}{n}\ \ \ }{\ \ \ \frac{g^{\alpha}-1 \!\!\!\mod  n^2}{n}\ \ \ }=\frac{sm}{s}=m \ \ \ \!\!\!\!\!\mod  n.$$

\textbf{Correctness of Variant 2}. The variant takes $g=1+n, x=m, y=r$ in Problem 2.
It is easy to find that
\begin{eqnarray*}
\frac{c^{\kappa}-1 \!\!\!\mod  n^2}{n}&=&\frac{((1+n)^m r^n)^{\tau\lambda}-1 \!\!\!\mod  n^2}{n}\\
&=&\frac{(1+n)^{m\tau\lambda}-1\!\!\!\mod  n^2}{n}\\
&=&\frac{nm\tau\lambda \!\!\!\mod  n^2}{n}=\frac{nm}{n}=m
\end{eqnarray*}

\textbf{Correctness of Variant 3}.  The variant takes $x=m, y=r$ in Problem 2.
It is easy to check that
\begin{eqnarray*}
\frac{c^{\lambda}-1 \!\!\!\mod  n^2}{n}&=&\frac{(g^m r^n)^{\lambda}-1 \!\!\!\mod  n^2}{n}\\
&=&\frac{(g^{\lambda})^m-1 \!\!\!\mod  n^2}{n}\\
&=&\frac{(1+n)^m-1 \!\!\!\mod  n^2}{n}=m
\end{eqnarray*}

\textbf{Correctness of Variant 4}. It is easy to see  that
\begin{eqnarray*}
\frac{\frac{c}{(c^d\!\!\!\mod  n)^e}-1\!\!\!\mod  n^2}{n}&=& \frac{\frac{(1+n)^m r^e}{(((1+n)^m r^e)^d\!\!\!\mod  n)^e}-1\!\!\!\mod  n^2}{n}\\
 &=&\frac{\frac{(1+n)^m r^e}{r^e}-1\!\!\!\mod  n^2}{n}=m
 \end{eqnarray*}

\subsection{The Bresson-Catalano-Pointcheval encryption scheme revisited}

 The Bresson-Catalano-Pointcheval encryption scheme  has not directly specified that $n\,|\,\mbox{ord}_{n^2}(g)$. But it is easy to find that such a picked $g$ satisfies the condition \underline{with high probability}.  In view of that the condition is necessary to recover $x$ in Problem 1 (see the proof of Theorem 1),   we shall directly specify it in the Setup phase.

  The random pad $r$ is chosen by the sender  and  is blinded as
  $$A=g^r \!\!\!\mod  n^2, \quad B=h^r(1+mn)\!\!\!\mod  n^2.$$
   We have
$$A=g^r\cdot 1^n \!\!\!\mod  n^2,  $$
here it takes $x=r, y=1$ in Problem 1. Thus one knowing the trapdoor $\lambda$ can recover $r$ using Paillier's computational method.
Note that although $B$ could be viewed as
$$B=(1+n)^mh^r\!\!\!\mod  n^2 $$
it does not fall into the class of Problem 1. One cannot recover $r$ from $B$ whether the trapdoor is known or not.

 After $r$ is retrieved, one can  recover  $m=  \frac{B/h^r-1 \!\!\!\mod  n^2}{n} $ directly.  Obviously, the original computational method incurs more cost.
 Based on the observation, we now present a revision of the scheme as follows.

\begin{center}
Table 5: The Bresson-Catalano-Pointcheval encryption scheme revisited

\begin{tabular}{|l|l|l|}
  \hline
  & $n=pq$,  $\lambda=\mbox{lcm}(p-1, q-1)$.   & $n=pq$,  $\lambda=\mbox{lcm}(p-1, q-1)$.  \\
  \multirow{3}*{\centering\parbox{0.45in} {Setup}} & $\alpha\in \mathbb{Z}_{n^2}^*$, $a<n\lambda/2$,  &   $g \in \mathbb{Z}_{n^2}^*$, $n\,|\,\mbox{ord}_{n^2}(g)$.    \\
 & $g=\alpha^2\!\!\!\mod  n^2, h=g^a\!\!\!\mod  n^2 $. & $a\in \mathbb{Z}_n^*$, $h=g^a\!\!\!\mod  n^2. $    \\
& $\rho=\left(\frac{g^{\lambda}-1 \!\!\!\mod  n^2}{n}\right)^{-1}\!\!\!\mod  n. $   & $\rho=\left(\frac{g^{\lambda}-1 \!\!\!\mod  n^2}{n}\right)^{-1}\!\!\!\mod  n. $   \\
&$\tau=\lambda^{-1}\!\!\!\mod  n.$ & \\
&PK: $n, g, h$; SK:   $a, \lambda, \rho, \tau$.  &  PK: $n, g, h$; SK:   $a, \lambda, \rho$.  \\    \hline
\multirow{4}*{\centering\parbox{0.45in} {Enc.}}  & For $m\in \mathbb{Z}_{n}$, pick $r\in \mathbb{Z}_{n}$,  & \multirow{4}*{\centering\parbox{2.1in} {It is the same as  the original.}}   \\
&compute $A=g^r \!\!\!\mod  n^2$, &  \\
 & $B=h^r(1+mn)\!\!\!\mod  n^2.$  &   \\
& The ciphertext is  $c=(A, B)$.  &    \\ \hline
Dec. 1   & $m=\frac{B/A^a-1 \!\!\!\mod  n^2}{n}$   & It is the same as  the original.  \\ \hline
\multirow{3}*{\centering\parbox{0.45in} {Dec. 2}} & $r= \rho\left(\frac{A^{\lambda}-1 \!\!\!\mod  n^2}{n}\right)\!\!\!\mod  n. $   &  $r= \rho\left(\frac{A^{\lambda}-1 \!\!\!\mod  n^2}{n}\right)\!\!\!\mod  n. $  \\
&$\gamma=ar\!\!\!\mod  n.$ & \\
& $m=\frac{\left(\frac{B}{g^{\gamma}}\right)^{\lambda}-1 \!\!\!\mod  n^2}{n}\cdot \tau\!\!\!\mod  n,$  &   $m=  \frac{B/h^r-1 \!\!\!\mod  n^2}{n} $ \\
 \hline
\end{tabular}\end{center}

We stress that the new alternative decryption method does not invoke the secret parameter $a$, which means the secret parameters $a, \lambda, \rho$ can be divided into two groups, $a$ and $(\lambda, \rho)$. The two groups of secret parameters can be allocated to different users so as to enhance the robustness of key management. The new version is more flexible  to be integrated into other cryptographic schemes.

\subsection{Comparisons}

The Paillier's encryption scheme can be naturally transformed into a signature scheme because it can retrieve the random pad $r$  as well as the message $m$. This is due to that it
only requires  $n\,|\,\mbox{ord}_{n^2}(g)$. But in the Variant 1, one cannot retrieve  $r\in \mathbb{Z}_n^*$,  instead $g^r\!\!\!\mod  n^2$. Though the Variant 3 is very similar to the original Paillier's encryption scheme, it is insecure against a chosen ciphertext attack \cite{ST02}. The others, Variant 2, Variant 4 and Variant 5 cannot be transformed into signature schemes. See the following Table 6 for details.

By the way, the claim that some variants are more efficient than the original Paillier's encryption scheme is somewhat misleading. Actually, in Paillier's encryption scheme the computation
$\left(\frac{g^{\lambda}-1 \!\!\!\mod  n^2}{n}\right)^{-1} \!\!\!\mod  n$
has no relation to the ciphertext $c$.  It can be computed and stored previously. The dominated computation in the decryption procedure is that
$\frac{c^{\lambda}-1 \!\!\!\mod  n^2}{n}$, while the corresponding computation in Variant 4 is
$m=\frac{\ \ \frac{c}{(c^d\!\!\!\mod  n)^e}-1\!\!\!\mod  n^2}{n},$
and that in Variant 5 is
$m=\frac{B/A^a-1 \!\!\!\mod  n^2}{n}.$
We find these decryptions require almost the same
 computational cost.

\begin{center}
Table 6:  Comparisons of Paillier's encryption and  Some variants

\begin{tabular}{|l|l|}
  \hline
  \multirow{3}*{\centering\parbox{0.8in} {The original}}   &   $c= g^mr^n\!\!\!\mod  n^2.$  $n\,|\,\mbox{ord}_{n^2}(g), x=m, y=r.$   \\
 &  \\
 & Verification  w.r.t. $(m, s_1, s_2)$:  $H(m)\stackrel{?}{=}g^{s_1}s_2^n\!\!\!\mod  n^2.$ \textbf{True}.
   \\   \cline{2-2}
   \hline \hline
  \multirow{4}*{\centering\parbox{0.8in} {Variant 1}}   &   $c= g^m(g^r)^n =g^{m+rn}\!\!\!\mod  n^2.$   \\
 &  $\mbox{ord}_{n^2}(g)=\alpha n, x=m,$ $y=g^r$ is a special random pad. \\
  & Verification  w.r.t. $(m, s_1, s_2)$: $H(m)\stackrel{?}{=}g^{s_1+s_2n}\!\!\!\mod  n^2$.  \textbf{False}.
    \\   \cline{2-2}
   \hline \hline
     \multirow{3}*{\centering\parbox{0.8in} {Variant 2}}  &   $c=(1+n)^m r^n=(1+mn) r^n\!\!\!\mod  n^2.$ \\
       &  $g=1+n, \mbox{ord}_{n^2}(g)=n,  x=m, y=r$ \\
        & Verification w.r.t. $(m, s_1, s_2)$: $H(m)\stackrel{?}{=}(1+s_1n) s_2^n\!\!\!\mod  n^2$. \textbf{False}.
    \\   \cline{2-2}
 \hline  \hline
  \multirow{1}*{\centering\parbox{0.8in} {Variant 3}}  &  $c=g^m r^n \!\!\!\mod  n^2.$
  $g^{\lambda}=1+n,  x=m, y=r$\\
  & It can \textbf{not} resist  a chosen
ciphertext attack.  \\  \hline \hline
 \multirow{3}*{\centering\parbox{0.8in} {Variant 4}}
   &    $c=(1+mn) r^e\!\!\!\mod  n^2,$ $g=1+n, ed=1 \!\!\!\mod  \phi(n)$.    \\
  &    \\
   & Verification  w.r.t. $(m, s_1, s_2)$:  $H(m)\stackrel{?}{=}(1+s_1n) s_2^e\!\!\!\mod  n^2$. \textbf{False}.
  \\   \cline{2-2}
    \hline \hline
 Variant 5   &  $(A, B)=(g^r \!\!\!\mod  n^2,  (1+mn)h^r\!\!\!\mod  n^2)$  \\
  (Bresson-Catalano  &  \\
    -Pointcheval)  & Verification  w.r.t. $(m, s_1, s_2)$: $H(m)\stackrel{?}{=}(1+s_1n)h^{s_2}\!\!\!\mod  n^2$. \textbf{False}.
    \\   \cline{2-2}
   \hline
\end{tabular}\end{center}

\section{Conclusion} We revisit the Paillier's cryptosystem and present an efficient alternative decryption procedure for Bresson-Catalano-Pointcheval encryption scheme.
We reaffirm that the original Paillier's encryption scheme has a special property that it naturally implies a signature scheme, while those variants miss this feature.



\begin{thebibliography}{4}
\renewcommand{\baselinestretch}{1.1}
  \renewcommand{\arraystretch}{.9}
  \normalsize \small \parskip 0mm

  \bibitem{BCH13} Boneh, D., Craigentry, C., Halevi, S,  Wang F., Wu, D.:  Private Database Queries Using Somewhat
Homomorphic Encryption. In: Jacobson, M., et al. (Eds.),  Proc. of ACNS 2013, LNCS, vol. 7954, pp. 102-118. Springer, Heidelberg (2013)

  \bibitem{BCP03}  Bresson, E., Catalano, D., Pointcheval, D.: A Simple Public-Key Cryptosystem
with a Double Trapdoor Decryption Mechanism and Its Applications.
In: Laih, C. (ed.), Proc. of ASIACRYPT 2003. LNCS, vol. 2894, pp. 37-54.
Springer, Heidelberg (2003)

\bibitem{CCW01} Choi, D., Choi, S., Won, D.: Improvement of Probabilistic Public
Key Cryptosystem Using Discrete Logarithm.  In: Kim, K. (ed.), Proc. of ICISC 2001, LNCS, vol. 2288,
pp.72-80. Springer, Heidelberg (2002)

\bibitem{CGH01} Catalano, D., Gennaro, R., Howgrave-Graham, N., Nguyen, P,: Paillier's
Cryptosystem Revisited, In: Proc. of ACM CCS 2001, pp. 206-214 (2001)

\bibitem{CL15} Castagnos, G., Laguillaumie, F.: Linearly Homomorphic Encryption from DDH.
 In: Nyberg, K. (ed.), Proc. of CT-RSA 2015, LNCS, vol. 9048, pp. 487-505. Springer, Heidelberg (2015)

 \bibitem{CMP14} Catalano, D.,  Marcedone, A., Puglisi, O.:  Authenticating Computation on Groups: New Homomorphic Primitives and Applications. In: Sarkar, P., Iwata T. (eds.),  Proc. of  ASIACRYPT 2014,  LNCS, vol. 8874, pp. 193-212. Springer, Heidelberg (2014)

\bibitem{DJ01}   Damg{\aa}rd, I., Jurik, J.: A Generalisation, a Simplification and Some
Applications of Paillier's Probabilistic Public-Key System. In: Kim, K. (ed.), Proc. of PKC 2001. LNCS, vol. 1992, pp. 119-136. Springer,
Heidelberg (2001)

\bibitem{DJN10} Damg{\aa}rd, I., Jurik M., Nielsen, J.:  A Generalization of Paillier's Public-key System with Applications
to Electronic Voting.  Int. J. Inf. Secur. 2010, 9: 371-385 (2010)

\bibitem{G02}  Galbraith, D.: Elliptic Curve Paillier Schemes. J. Cryptology 15(2), 129-138 (2002)

\bibitem{G09} Gentry, C.: Fully Homomorphic Encryption Using Ideal Lattices. In: Proc. of
STOC 2009, pp. 169-178. ACM (2009)

\bibitem{GM84} Goldwasser, S., Micali, S.: Probabilistic Encryption. JCSS 28(2), 270¨C299 (1984)

\bibitem{JL13} Joye, M., Libert, B.: Efficient Cryptosystems from 2k-th Power Residue
Symbols. In: Johansson, T., Nguyen, P.Q. (eds.), Proc. of EUROCRYPT 2013. LNCS, vol. 7881, pp. 76-92. Springer, Heidelberg (2013)

\bibitem{NS98} Naccache, D., Stern, J.: A New Public Key Cryptosystem Based on Higher
Residues. In: Proc. of ACM CCS 1998, pp. 546¨C560 (1998)

\bibitem{OU98} Okamoto, T., Uchiyama, S.: A New Public-Key Cryptosystem as Secure
as Factoring. In: Nyberg, K. (ed.), Proc. of EUROCRYPT 1998. LNCS, vol. 1403,  pp. 308-318. Springer, Heidelberg (1998)

\bibitem{P99} Paillier, P.: Public-Key Cryptosystems Based on Composite Degree Residuosity
Classes. In: Stern, J. (ed.), Proc. of EUROCRYPT 1999. LNCS, vol. 1592,p
pp. 223-238. Springer, Heidelberg (1999)

\bibitem{ST02} Sakurai, K.,  Takagi, T.:   On the Security of a Modified Paillier
Public-Key Primitive, In:  Batten, L. and Seberry, J. (eds.), Proc. of  ACISP 2002, LNCS, vol. 2384, pp. 436-448.  Springer, Heidelberg (2002)

\bibitem{ST06} Schoenmakers, B.,  Tuyls, P.: Efficient Binary Conversion for Paillier Encrypted Values. In: Vaudenay, S. (ed.),
 Proc. of EUROCRYPT 2006, LNCS, vol. 4004, pp. 522-537. Springer, Heidelberg (2006)

\end{thebibliography}
\end{document}